\documentclass[preprint,preprintnumbers,amsmath,amssymb]{revtex4}
\usepackage{graphicx}
\usepackage{dcolumn}
\usepackage{bm}
\usepackage{color}

\begin{document}
\title{Statistical method for A-RNA and B-DNA 
}

\author{Marco Zoli}

\affiliation{School of Science and Technology \\  University of Camerino, I-62032 Camerino, Italy \\ marco.zoli@unicam.it \\ ORCID / 0000-0003-1739-0757}

\date{\today}

\begin{abstract}
Nucleic acids have been regarded as stiff polymers with long-range flexibility and generally modeled using elastic rod models of polymer physics. Notwithstanding, investigations carried out over the past few years on single fragments of order $\sim 100$ base pairs have revealed remarkable flexibility properties at short scales and called for theoretical approaches that emphasize the role of the bending fluctuations at single sites along the molecule stack. 
Here, we review a three dimensional mesoscopic Hamiltonian model which assumes a discrete representation of the double stranded (ds) molecules at the level of the nucleotides. The model captures the fundamental local interactions between adjacent sugar-phosphate groups and the pairwise interactions between complementary base pair mates.  
A statistical method based on the path integral formalism sets the ensemble of the base pair breathing fluctuations which are included in the partition function and permits to derive the thermodynamics and the elastic response of single molecules to external forces.
We apply the model to the computation of the twist-stretch relations for fragments of ds-DNA and ds-RNA, showing that the obtained opposite pattern (DNA overtwists whereas RNA untwists versus force) follows from the different structural features of the two helices. Moreover, we focus on the DNA stretching due to the confinement in nano-pores and, finally, on the computation of the cyclization probability of open ends molecules of $\sim 100$ base pairs under physiological conditions. The mesoscopic model shows a distinct advantage over the elastic rod model in estimating the molecule bendability at short length scale.
 

\end{abstract}

\maketitle
\tableofcontents
\newpage

\section*{1.  \, Introduction}

Nucleic acids are polymeric chains whose building blocks are the nucleotides. Each nucleotide is composed of a sugar ring, a phosphate group and a nitrogenous base. The bases are connected by (glycosidic) covalent bonds to the deoxyribose sugar in DNA and to the ribose sugar in RNA. 
The sugar of one nucleotide is connected to the phosphate group of the adjacent nucleotide by strong (phosphodiester) covalent bonds which impart directionality to the polynucleotides strand. Nitrogen bases on complementary strands form a base pair whose specificity is provided by the Watson-Crick pairing. The base pairs are held together by hydrogen bonds which are weaker than covalent bonds thus allowing for the breathing fluctuations of the double stranded helical molecules \cite{calla}.
The two complementary strands make the molecule backbone. As each nucleotide is made of at least $50$ atoms, fully atomistic models for nucleic acids are computationally challenging and limited to short oligomers in molecular dynamics simulations. 

In vivo, the polynucleotide chain is subject to thermal fluctuations that knock the molecular bonds and uncorrelate the orientations of its repeated units thus lending support to representations of nucleic acids in terms of random coils.
Mathematical models of polymer physics have been applied for many years to study the elastic properties of DNA (by far more investigated than RNA), since it was recognized that the molecule flexibility is essential e.g., to its compaction into eukaryotic nucleosomes and to its biological functioning, including the binding of proteins which regulate transcriptional activity \cite{maher10}. 

In the simplest model, the Freely Jointed Chain, DNA is treated as a chain of $N$ straight segments of equal length $l_K$ linked by joints around which they can rotate freely. As there is no correlation between the directions of different bond vectors, the mean end-to-end distance is zero and the average square end-to-end distance is $N\cdot l_{K}^{2}$, the signatures of a random walk \cite{flory}. While the FJC neglects the excluded volume effects for a polymer in a solvent (self avoiding walks would swell the polymer size), it has the merit to capture the entropic origin of the polymer elasticity as it was first seen in extension versus force experiments on single molecules of lambda phage DNA \cite{busta92}. If the applied stretching forces are small, the thermal bending fluctuations prevail and cause the helix to assume different random walk configurations.  At room temperature, this occurs for forces below $\sim 0.04 pN$, as this is the ratio of $k_BT$ to the Kuhn length  $l_K \sim 100 nm$, whereby the latter defines the length scale over which the polymer is rigid. In this regime, the FJC can describe entropic elasticity and the extension of $\lambda $-DNA  remains smaller than $\sim 0.3 L_c$, with $L_c \sim 16\mu m$ being the contour length of the molecule.  

In order to stretch the molecule between $\sim 0.3 L_c$ and $\sim 0.9 L_c$, increasing forces (larger than $\sim 0.04 pN$ and up to $\sim 10 pN$) should be applied \cite{busta94}. This stems from the fact that more work is required to suppress thermal bending fluctuations at successively shorter length scales. Here, the FJC model fails to fit the data as the molecule elastic response becomes non linear and the cooperative coupling between neighboring bonds (neglected by the FJC model) takes over. Instead, in this low to intermediate force range, the inextensible Worm-Like-Chain model  well describes the DNA elasticity. The WLC model, also termed as the continuous limit of the Kratki-Porod model \cite{kp,kp1}, conceives the molecule as a long continuous rod that bends smoothly due to thermal fluctuations. As a result, the infinitesimally small chain bonds maintain an orientational correlation which decays exponentially over a characteristic persistence length $l_p$, a measure of the molecule stiffness. The stiffer the molecule, the longer $l_p$. Note that the FJC model represents a reduced form of the WLC model whereby the stiffness of the discretized polymer, i.e. the Kuhn length, is twice the standard $l_p$.
For ds-DNA under physiological conditions ($10 \, mM$ salt buffer), $l_p$ is $\sim 50 nm$  although a significant softening may occur at higher ionic strengths \cite{save12}. Conversely, in a low salt buffer, long range electrostatic effects increase the effective $l_p$ of the charged polymer chain and may make it dependent on the stretching force scale \cite{odj77,fix77,barrat93,marko95,croq99}.

At forces larger than $\sim 10 pN$, the molecule end-to-end distance slightly exceeds the contour length $L_c$ showing a deviation from the predictions of the inextensible WLC model. This follows from the elongation of the chain bonds that cause an intrinsic stretching of the molecule and marks the transition to a regime in which enthalpic contributions are relevant. In this range the WLC model can be modified to incorporate the enthalpic stretching and account for the elastic response of the molecule  \cite{odj95,block97}. 

Finally, at forces $\sim 65 pN$, the molecule undergoes an abrupt elongation from $\sim 1.05 L_c$ to $\sim 1.7 L_c$ known as the overstretching transition. The latter has been interpreted as a conversion of double stranded to single stranded DNA \cite{mameren09} although the details of the transition may depend on the sequence specificity and also on the attachment geometry that torsionally constrains the molecule strands in optical tweezers experiments \cite{busta12}. In vivo, DNA overstretching is known to occur as a consequence of proteins binding to the chain in recombination and repair processes \cite{stasi97,marko98,ha12}. 

To sum up, the WLC model yields a good representation of the force-extension relations as long as the molecule extension, measured by the end-to-end distance, remains  $\lesssim  1.05 L_c$ and, more generally, the WLC model accounts for the DNA flexibility properties provided that $L_c$ is larger than $l_p$, i.e. at length scales of at least $\sim 150$ base pairs. For shorter oligomers, recent extensions of the model which include twisting and bending deformations (TWLC) predict scale dependent persistence lengths, with a soft behavior at scales  $\lesssim  20$ base pairs  both for DNA and RNA, if non local couplings between distant sites (nlTWLC) are taken into account \cite{carlon21,carlon22,sakaue23}.

Experimentally, small-angle x-ray scattering  (SAXS) data for the radius of gyration together with fluorescence resonance energy transfer (FRET) analysis of the end-to-end distance distribution   \cite{archer08},  have put forward the idea that open ends fragments of $\sim 100$ base pairs may have a persistence length smaller than the above mentioned standard value for ds-DNA in the kilo base pair range. 

{Moreover, it was shown long ago, by analysis of the cyclization probability induced by T4 ligase enzymes on linear DNA fragments \cite{shore81}, that the looping probability steadily decreases for fragments smaller than $\sim 500$ base pairs due to the fact that the energetic cost required to meet the chain ends is progressively higher.  Nevertheless, ds-DNA chains still display an appreciable flexibility also for fragments whose length is of order $l_{p}$  \cite{horo}. 
}
{Successive experiments based on ligase dependent cyclization assay \cite{widom} and  FRET  \cite{vafa,kim13}, have derived cyclization probabilities ($J$-factors) much larger than the values obtained by the WLC model  for fragments of $\sim 100$ base pairs as it will be shown in Section 4.C. This points to the fact that DNA molecules display an intrinsic flexibility also at those short length scales involved e.g., in the DNA wrapping around histone proteins and compaction into nucleosomes. 
}

{ All these findings, stemming from different techniques, consistently support the view that the DNA mechanical properties may depend on the fragment size \cite{gole12} and challenge the applicability of the standard WLC model to investigate the DNA elastic properties at short length scales \cite{maiti15}.
}

For these reasons, analysis of the flexibility properties and, in particular, of the $J$-factors in short molecules have been carried out more recently by using
mesoscopic Hamiltonian models as a viable alternative to the continuous WLC model \cite{io16a,io17}. Mesoscopic models have the advantage to treat DNA  at the level of the base pair, thus allowing for broad radial fluctuations and large bending angles between adjacent nucleotides that, in turn, favor the formation of kinks which locally unstack the helix \cite{volo08,zocchi13,kim14,ejte15}, enhance the cyclization efficiency and may reduce the persistence length  \cite{tan15,io16b,onuf19}.

In general, coarse-grained Hamiltonian models  that assume a point-like representation for a single nucleotide  are widely applied to calculate the structural, thermodynamical and transport properties of nucleic acids \cite{io11b,hando12,io14c,albu14,singh15,weber15,io23,kavitha25}. Studies of melting profiles, base pair breathing fluctuations and bubble statistics, mainly of DNA,
are usually based either on a Poland-Scheraga model that conceives the molecule as a Ising-like chain of open/closed states \cite{pol66} or on a Peyrard-Bishop (PB) model first introduced to account for the DNA melting transition \cite{pey89} and later modified (PBD model) to incorporate non linear effects in the stacking potential \cite{pey93}. Recently, also an extended version of the PBD model has been proposed to account for bubble lifetimes and lengths distributions in heterogeneous DNA sequences by means of molecular dynamics simulations \cite{kalos20}.

Focusing on the the base pair dynamics, the PBD model has the advantage to describe both  the intra-strand forces between adjacent bases along the stack and the hydrogen bonds between pair mates in terms of a single continuous variable, i.e., the base pair stretching mode.
This property is essential in order to treat the intermediate base pair states and the molecule dynamics \cite{campa98,singh16,lak19}.  

Notwithstanding this merit, it should be recognized that the original PBD model represents a DNA chain in the thermodynamic limit and in an infinitely diluted solution. 
Precisely, taking a Morse potential to represent the base pairs hydrogen bonds,  it follows that the distance between the pair mates can become arbitrarily large on the Morse plateau whenever the thermal fluctuations energy is sufficiently high to overcome the base pair binding energy and break the hydrogen bonds.

Clearly this picture is at odds with experiments in which the DNA concentrations in solutions are finite hence, single strands may recombine or single bases may form hydrogen bonds with the solvent once broken pairs get out of the stack \cite{io12,mukher24}.  Therefore, in the real cases,  the DNA bases always fluctuate in a finite environment and, accordingly, all computational methods  have to operate a restriction of the phase space available to the base pair displacements. From a theoretical viewpoint, the upper bound on the base pair distance is also required on general grounds in order to avoid the divergence of the partition function which stems from the fact that the on site Morse potential is bound \textit{and} breaks the translational invariance of the system \cite{zhang97,io16b}.

In this article we focus on these issues considering a mesoscopic Hamiltonian model for double stranded (ds)-DNA that goes beyond the PBD model and accounts for the three dimensional nature of the helix since it incorporates the twisting and  bending degrees of freedom between adjacent base pairs along the molecule stack. First, we describe
the general statistical framework, based on the path integrals computational method,  that allows one to consistently determine the cutoff on the amplitude of the base pair fluctuations and then we extend the model
to deal with ds-RNA whose helical features substantially differ from those of ds-DNA. As an application, we  focus on the interplay between structure and function,  arguing that the opposite twist-stretch behavior of ds-DNA and ds-RNA can be ascribed to their different helical forms \cite{bohr11,herrero17}.
Next, we show how the Hamiltonian model can be modified to study the stretching properties of DNA fragments in a cylindrical channel, given the growing relevance of this confinement method for the purpose of accurate single molecule sequencing and genome mapping. Finally, we present the calculation of the cyclization probability for short homogeneous DNA fragments comparing the results with the predictions of the WLC model and the available experimental data.

The 3D mesoscopic Hamiltonian model is described in Section 2 while the method which sets the fluctuations cutoff is presented in Section 3. The obtained results for the elastic properties of short fragments of nucleic acids are reported in Section 4 while  some final remarks are made in Section 5.

\section*{2. \, Model}

\renewcommand{\theequation}{2.\arabic{equation}}
\setcounter{equation}{0}

\begin{figure}
\includegraphics[height=8.0cm,width=8.0cm,angle=-90]{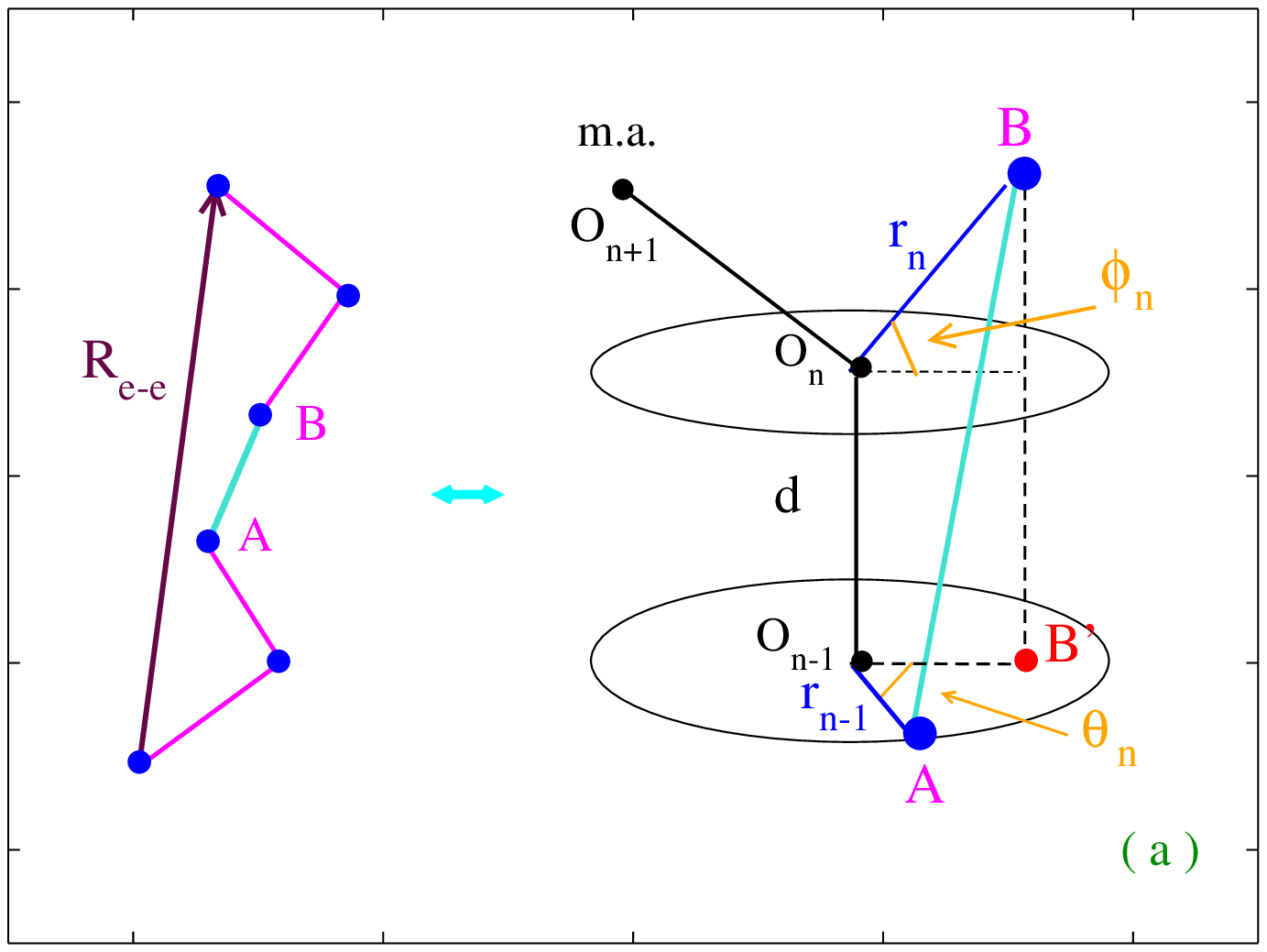}
\includegraphics[height=8.0cm,width=8.0cm,angle=-90]{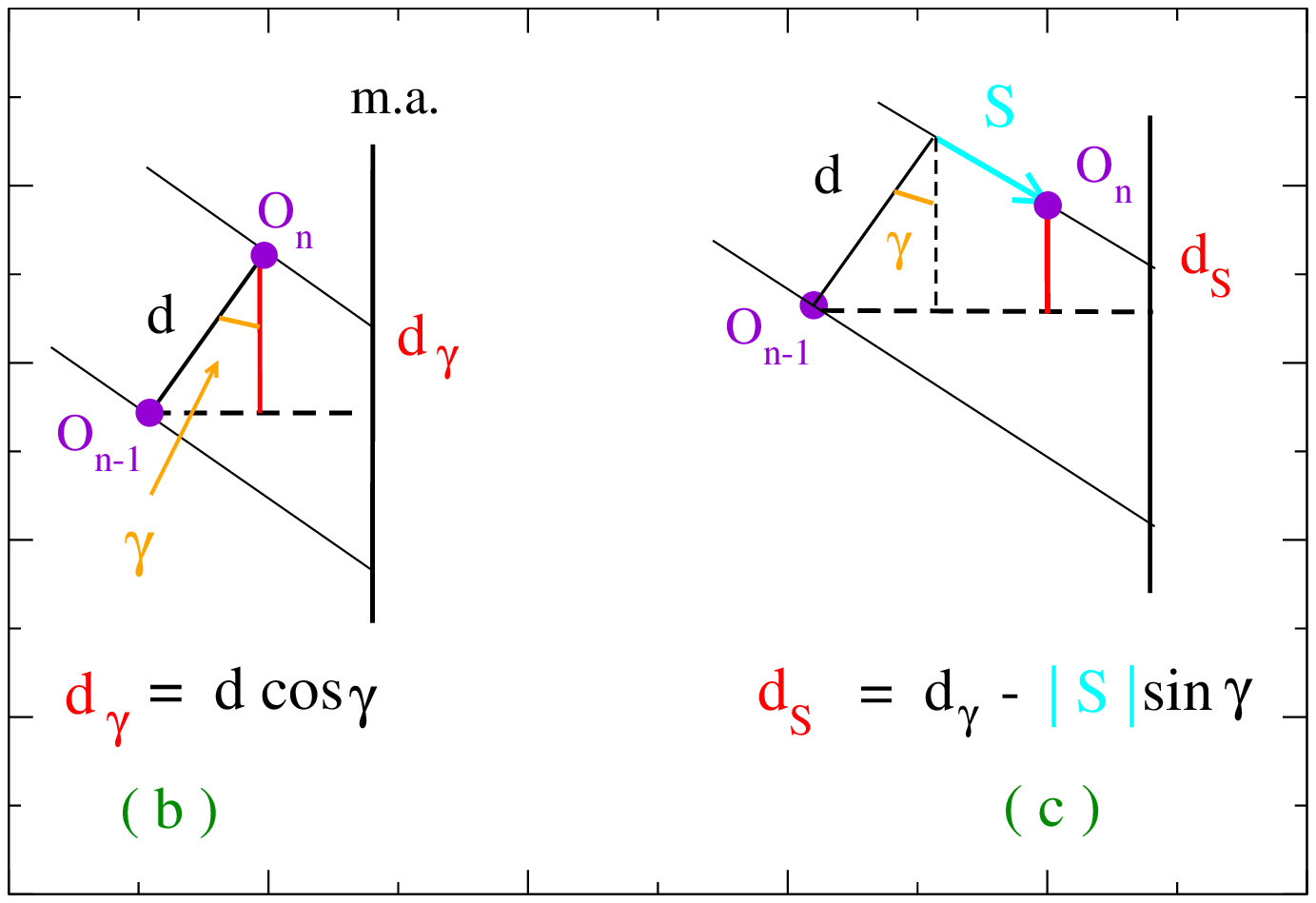}
\caption{\label{fig:1}(Color online)  
Reprinted  from ref.\cite{io23}. (a) 3D model for a helical chain.  $\phi_{n}$ is the bending angle between adjacent $r_{n}$'s in a dimer;  $\theta_n$ is the twist accumulated along the helix.  $\overline{AB}$ is the average dimer distance. $d$ is the rise distance between successive $O_{n}'s$ arranged along the mid helical axis.
$R_{e-e}$ is the end-to-end distance. (b) Base pair inclination with respect to the helical axis measured by the  tilt angle $\gamma $: the rise distance (along the molecular axis) $d_{\gamma }$ is shorter than $d$.
(c) The n-th base pair slides on top of the adjacent (n-1)-th base pair.  $S$ is taken as negative hence, the rise distance $d_{S}$ gets shorter than $d_{\gamma }$. 
}
\end{figure}

Several studies produced over the last years have investigated denaturation bubbles, thermodynamics, flexibility, force/stretching relations and length distribution functions of linear chains and loops with variable size and sequence \cite{io14b,io18,io18a,io19}. These works are based on a three dimensional model, whose schematic in shown Fig.~\ref{fig:1}(a),  in which the $r_{n}$'s measure the radial fluctuating distances between the complementary base pair mates with respect to the mid helical axis. When the fluctuations vanish, i.e. the blue dots overlap the respective $O_n$'s arranged along the mid axis, then the pair mates distances equal the average helix diameter $R_0$. 

Clearly, this point-like representation provides a coarse-grained description of the helix and neglects those degrees of freedom, i.e. the rotations between the two bases of a pair, which are instead taken into account in models depicting the base pairs as rigid blocks \cite{calla,dick89} 

Nevertheless, the model has sufficient structure to deal with rotations between neighboring base pairs along the molecule stack. Specifically, for a given dimer, adjacent base pairs are locally bent by an angle $\phi_{n}$ and twisted by $\theta_n$.
If the average bending angles are close to zero, then
the $n-th$ base pair radial fluctuations occur in a plane essentially perpendicular to the molecular axis \cite{io11b}. If this feature is common to a large number of dimers in the chain then there is no overall significant tilt of the base pairs planes. This picture is assumed to model the physiological B-form of ds-DNA, in which the helix axis runs through the center of each base pair and the pairs are stacked almost per­pendicular to the axis. 

However, the latter condition is not fulfilled in the  A-form helix that displays two microscopic features, visualized in Fig.~\ref{fig:1}(b), whose entity is dependent on the di-nucleotide step:  i) the base pairs planes are tilted by the angle $\gamma $ respect to the vertical helical axis and ii) the base pairs forming a dimer, slide by a distance $S$ past each other. As a result, the A-form helix is more compact and broader than the B-form. For simplicity it is hereafter assumed that, for a single simulation, $\gamma $ and $S$ are average values distinctive of the chain although local variations are found in heterogeneous sequences according to the specific base pair steps.
Altogether, the occurrence of tilt and slide has the direct consequence to shorten the rise distance along the helical axis, a feature that may affect the stretching flexibility. In the following we will take the A-form helix as appropriate for ds-RNA while the B-form (with zero $\gamma $ and $S$) is assumed suitable for ds-DNA although it has to be mentioned that irregularities in the di-nucleotide steps with significant slide at specific sites are found also in B-DNA structures and are believed to be instrumental to the helix packaging into nucleosomes \cite{olson07}.

The chain depicted in Fig.~\ref{fig:1}(a) is treated by a Hamiltonian containing the main forces that stabilize the double helix. Explicitly, for a system of $N$ base pairs, it reads:

\begin{eqnarray}
& &H =\, H_a[r_1] + \sum_{n=2}^{N} H_b[r_n, r_{n-1}, \phi_n, \theta_n] \, , \nonumber
\\
& &H_a[r_1] =\, \frac{\mu}{2} \dot{r}_1^2 + V_{1}[r_1] \, , \nonumber
\\
& &H_b[r_n, r_{n-1}, \phi_n, \theta_n]= \,  \frac{\mu}{2} \dot{r}_i^2 + V_{1}[r_n] + V_{2}[ r_n, r_{n-1}, \phi_n, \theta_n]  \, \, , \nonumber
\\ 
& &V_{1}[r_n]=\, D_n \bigl[\exp(-b_n (|r_n| - R_0)) - 1 \bigr]^2  \, , \nonumber
\\
& &V_{2}[ r_n, r_{n-1}, \phi_n, \theta_n]=\, K_{n, n-1} \cdot \bigl(1 + G_{n, n-1}\bigr) \cdot \overline{d_{n,n-1}}^2   \, , \nonumber
\\
& &G_{n, n-1}= \, \rho_{n, n-1}\exp\bigl[-\alpha_{n, n-1}(|r_n| + |r_{n-1}| - 2R_0)\bigr]  \, . 
\label{eq:01}
\end{eqnarray}

$H_a[r_1]$ refers to the first base pair in the sequence  which lacks the preceeding base pair along the chain. 

$V_{1}[r_n]$ represents the one-particle inter-strands potential and $V_{2}[ r_n, r_{n-1}, \phi_n, \theta_n]$ is the two-particles intra-strand stacking potential.

{$V_{1}[r_n]$ is usually modeled by a Morse potential that has a stable minimum for the equilibrium base pair separation, a plateau for large inter-strand separation and a hard core which accounts for the repulsive interaction between phosphate groups on complementary strands. $D_n$ is the base pair dissociation energy which sets the height of the Morse plateau while $b_n^{-1}$ measures the one-particle potential range. Due to the spatial fluctuations of the nucleotides, the pair mates separation may get smaller than the average helix diameter. However, the base pair mates cannot get too close to each other as these fluctuations would have a large energetic cost hence, they would yield a small statistical contribution to the partition function due to the Morse potential hard core. 
Thus, in order to select a subset of possible base pair fluctuations, we set the condition \, $V_{1}[r_n] \leq D_n$ \, and use in the numerical program only those radial displacements such that,  $|r_n| - R_0 \geq - \ln 2/b_n$. }

{Following the critical remarks presented in the Introduction, we modify $V_{1}[r_n]$ by introducing a term that mimicks the stabilizing effects of the solvent surrounding the molecule. Specifically, the new potential term increases the base pair dissociation energy and accounts for possible recombination processes between solvent and single stranded molecules.
For these purposes, we take a widely used expression which produces a barrier over the Morse plateau \cite{coll95,druk01,io11b}:
}

\begin{eqnarray}
V_{Sol}[r_n]=\, - D_n f_s \bigl(\tanh((|r_n| - R_0)/ l_s) - 1 \bigr)
\label{eq:01a}
\end{eqnarray}

{whereby the dimensionless $f_s$ is the factor that accounts for the increase of the energy barrier while the length $l_s$ measures the spatial range of the solvent.
}
{The intra-strand stacking potential $V_{2}[ r_n, r_{n-1}, \phi_n, \theta_n]$ contains the angular degrees of freedom. By suppressing the latter, one would recover the 1D non linear potential used in the PBD  model. It had been shown that a non linear stacking between adjacent base pairs could account for the cooperative formation of melting bubbles and explain the sharp thermal denaturation of DNA molecules, in the thermodynamic limit  \cite{pey93}. 
The dependence on the angular variables is contained in $\overline{d_{n,n-1}}$, that is the average distance $\overline{AB}$ in Fig.~\ref{fig:1}(a). Moreover, for each dimer, the stacking depends on three force constants \, i.e., the harmonic $K_{n, n-1}$ and the anharmonic  $\rho_{n, n-1}$, $\alpha_{n, n-1}$. Enforcing the condition \,
$\alpha_{n, n-1} < b_n$, one assumes that the two particle potential has a broader range than the Morse potential.} 

{Accordingly, if the condition \, $|r_n| - R_0 \gg \alpha_{n, n-1}^{-1}$ \, is fulfilled, then the base pair breaks and the local stacking force constant decreases from  $ \sim K_{n, n-1}(1 + \rho_{n, n-1})$ to $ \sim  K_{n, n-1}$  thus loosening the stacking and favoring the breaking also of the neighboring base pair. This mechanism simulates the cooperative formation of local bubbles that may further extend along the stack and drive the melting transition. 
}
Summing up, Eq.~(\ref{eq:01}) models the key forces at play in the helix through five site dependent parameters which can be fitted to thermodynamical and elastic data for specific sequences \cite{owcz04,eijck11,io20}. For homogeneous fragments, the site dependence of the model parameters is dropped.
{For short linear fragments, finite size effects are generally large and are incorporated by taking open boundary conditions together with a specific  parametrization for the terminal base pairs which accounts for weaker stacking interactions  \cite{zgarb14}.
}
It is pointed out that the Hamiltonian in Eq.~(\ref{eq:01}) is more structured than the 1D PBD Hamiltonian. In fact, as the latter lacks the twist degree of freedom, both adjacent base pair may fluctuate independently in the absence of a restoring force and yield a sizeable contribution to the partition function.  Accordingly, the stacking potential of the 1D PBD model contains large (and uncorrelated) base pair fluctuations. 

{Instead, the twist angle between the base pairs in a dimer provides a restoring force that stabilizes the stacking potential \cite{io12}: as an example, if a base pair fluctuation $r_n$ gets large, the neighboring base pair can only assume fluctuation values $r_{n-1}$ such that the energy scale of the potential $V_{2}$ remains of order $D_n$ or below. These pairs of correlated fluctuations mostly contribute to the partition function whereas those fluctuations such that $V_{2} \gg D_n$ have scarce statistical weight. For these reasons, the twisted stacking $V_{2}$ in Eq.~(\ref{eq:01}) contains cooperative forces between neighboring nucleotides that confer stability to the double stranded molecule \cite{ashwood23}.   
It is also known that a Hamiltonian model with a twist angle between adjacent base pair planes removes the instabilities of the 1D untwisted model \cite{zhang97,io11b}. }

Furthermore, the inclusion of the bending degree of freedom in Eq.~(\ref{eq:01}) permits to study the cyclization and flexibility properties of open ends chains and also to deal with the properties of circular DNA molecules which cannot be addressed by a 1D ladder model \cite{io13}. 

{In general,  to compute the thermodynamics and mechanical properties for the Hamiltonian model in Eq.~(\ref{eq:01}), one has to carry out integrals over the radial and the angular variables for all nucleotides in the molecule with a proper choice for the integration cutoffs. 
In particular, in order to devise a method that allows one to set the cutoff on the radial fluctuations, the specific dependence on the twist and bending angles of the single base pairs plays a minor role. Then, the angular degrees of freedom are substituted by average values $\bar \phi$ and $\bar \theta$ that can be tweaked as model parameters thus reducing the simulation time. Accordingly, the 3D nature of the model is maintained while one can study the radial cutoff as a function of the helical conformation.
}
{ 
It is also noticed that such dependence may be significantly altered in helices with a sizeable sliding motion (see Fig.~\ref{fig:1}(b)). 
In the following, the twist conformation of the molecule will be macroscopically measured by the helical repeat i.e., the average number of base pair per helix turn. 
}

\section*{3.  \, Method}

\renewcommand{\theequation}{3.\arabic{equation}}
\setcounter{equation}{0}

The model, for a chain with $N$ base pairs, is studied by the path integral computational method, discussed in a number of papers, see e.g. refs.\cite{io09,io10,io11,io14a}. 
{In the finite temperature path integral formalism, the base pair displacements $r_n$'s are treated as dynamical quantities whose time evolution is defined by functions $r_n(\tau)$ whereby  $\tau \in [0, \beta]$ is the Euclidean time and  $\beta$ is  the inverse temperature \cite{io97,io03}. Using the boundary condition along the time axis i.e., $r_n(0)=\,r_n(\beta)$, one can write $r_n(\tau)$ in terms of a Fourier series as follows:
}

\begin{eqnarray}
r_n(\tau)=\, R_0 + \sum_{m=1}^{\infty}\Bigl[(a_m)_n \cos( \frac{2 m \pi}{\beta} \tau ) + (b_m)_n \sin(\frac{2 m \pi}{\beta} \tau ) \Bigr] \,.
\label{eq:02a}
\end{eqnarray}

While the Fourier coefficients define in principle all possible choices of fluctuations for any base pair, the path integral calculation incorporates in the statistical partition function a subset of $r_n$'s which fulfill the physical requirements of  the model potential discussed in the previous Section. 
Moreover, for computational purposes one has to truncate consistently the integration range for the Fourier coefficients in Eq.~(\ref{eq:02a}). The procedure is outlined hereafter while more details are found in ref.\cite{io21}.

\subsection*{3.A  \, Fluctuations Cutoff}

Let's define $\bar U$ the dimensionless integral cutoff for the Fourier coefficients in Eq.~(\ref{eq:02a}). We focus, for instance, on the $n-1$th radial fluctuations marked by the blue dot A in Fig.~\ref{fig:1}(a) and set $j \equiv \, n-1$. From Eq.~(\ref{eq:02a}), noticing that the Fourier coefficients are integrated on a even domain (see below), one derives the initial condition ($\tau =\,0$) for the average pair mates separation, that is \, $< r_j(0) > =\,R_0$. At any later time $t$, $r_j$ may fluctuate around $R_0$ consistently with the physical properties of the Morse potential (that models the hydrogen bonds between the pair mates) and of the solvent potential (that models the environment surrounding the helical molecule). Accordingly, as the $r_j$ fluctuation is conceived as a constrained Brownian motion \cite{maj05}, we define $P_j(R_0,\, t)$ as the probability  that $r_j$ does not return to $R_0$ until $t$ whereby, for a specific $t$,  $P_j(R_0,\, t)$ is expressed as a sum over the particle histories $r_j(\tau)$ in the time range $[0, t]$:

\begin{eqnarray}
& &P_j(R_0,\, t)=\,   \oint Dr_{1} \exp \bigl[- A_a[r_1] \bigr] 
\cdot \prod_{n=2, \, n\neq j}^{N} \oint Dr_{n}  \exp \bigl[- A_b [r_n, r_{n-1}, \bar \phi, \bar \theta ] \bigr]  \cdot \,  \nonumber 
\\
& & 
\int_{r_j(0)}^{r_j(t)} Dr_{j} \exp \bigl[- A_b[r_j, r_{j-1}, \bar \phi, \bar \theta] \bigr]  \cdot
\prod_{\tau=\,0}^{t}\Theta\bigl[r_j(\tau) - R_0\bigr] \, ,
\label{eq:03aa}
\end{eqnarray}

It follows that two time variables, $t$ and $\tau$, are required to compute Eq.~(\ref{eq:03aa}).

$A_a[r_1]$ and $A_b[..]$ are the action functionals related to the Hamiltonian in Eq.~(\ref{eq:01}) by the expressions:

\begin{eqnarray}
& &A_a[r_1]=\,\int_{0}^{\beta} d\tau H_a[r_1(\tau)] \, , \nonumber 
\\
& &A_b [r_n, r_{n-1}, \bar \phi, \bar \theta]=\, \int_{0}^{\beta} d\tau H_b[r_n(\tau), r_{n-1}(\tau), \bar \phi, \bar \theta] \, , \nonumber 
\\
& &A_b [r_j, r_{j-1}, \bar \phi, \bar \theta]=\, \int_{0}^{t} d\tau H_b[r_j(\tau), r_{j-1}(\tau), \bar \phi, \bar \theta] \, ,
\label{eq:03}
\end{eqnarray}

{Note that $r_j(\tau)$ describes an open end trajectory ($\tau \leq t < \beta$) whose measure of integration is ($\int Dr_{j}$), while all other $r_n(\tau)$ fluctuations are closed along the time axis, $r_n(0) =\, r_n(\beta)$, and therefore their measure of integration  is $\oint Dr_{n}$. It is also important to emphasize that the two measures are intertwined as the $n-th$ and $j-th$ sites are interacting through the two particle potential. 
}
Consistently with the path expansion in Eq.~(\ref{eq:02a}), $\oint Dr_{n}$ is  explicitly given by:

\begin{eqnarray}
& &\oint {D}r_{n} \equiv  \prod_{m=1}^{\infty}\Bigl( \frac{m \pi}{\lambda_{cl}} \Bigr)^2  \times \int_{-\Lambda_{n}(T)}^{\Lambda_{n}(T)} d(a_m)_n \int_{-\Lambda_{n}(T)}^{\Lambda_{n}(T)} d(b_m)_n \, , \, 
\label{eq:03b}
\end{eqnarray}

where  $\Lambda_{n}(T)$ is the temperature dependent cutoff for the Fourier coefficients and $\lambda_{cl}$ is the classical thermal wavelength \cite{io11,io14a} . 

The Heaviside function $\Theta[..]$ selects those trajectories $r_j(\tau)$ which stay larger than the helix diameter for any evaluated $\tau$ in the integration range. Then, the program retains only those sets of Fourier coefficients fulfilling this constraint and adds their contribution to $P_j(R_0,\, t)$. 

\begin{figure}
\includegraphics[height=8.0cm,width=8.0cm,angle=-90]{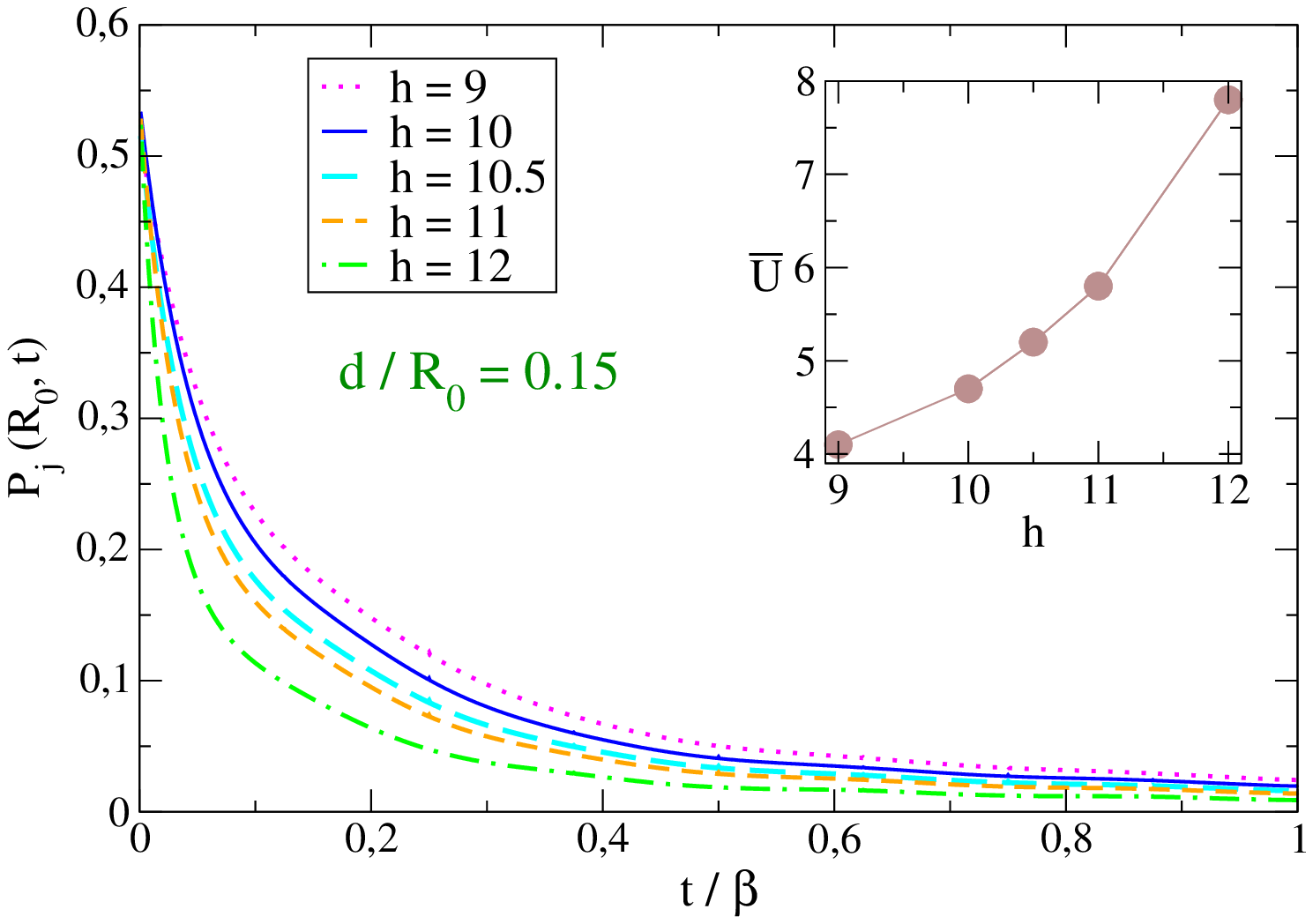}
\caption{\label{fig:2}(Color online) Reprinted  from ref.\cite{io21}. First-passage probability versus time for the $j-th$ base pair of the helical chain. The calculation is carried out at room temperature.  Five twist conformations are considered, each defined by a $h$ value.  For any helical repeat, the probability in Eq.~(\ref{eq:03aa}) is calculated by taking the associated cutoff \={U} displayed in the inset. 
}
\end{figure}

Let's now state the criterion to calculate $\bar U$. From the path expansion in Eq.~(\ref{eq:02a}), one observes that, at $t=\,0$, the $j-th$ trajectory is \, $r_j(0)=\, R_0 + \sum_{m=1}^{\infty}(a_m)_j$. As the Fourier coefficients are integrated on an even domain, the $P_j(R_0,\, 0)$ value obtained from Eq.~(\ref{eq:03aa}) should be $ \sim 1/2$ whereby the approximation sign stems from the constraint that large negative $(a_m)_j$ coefficients should be discarded since they are not allowed by the hard core of the Morse potential. Then, the zero time probability to have a radial amplitude larger than $R_0$  provides the benchmark to select the meaningful, time dependent probabilities as a function of the cutoff.
Technically, the integration $\int_{r_j(0)}^{r_j(t)} Dr_{j}$  in Eq.~(\ref{eq:03aa}) is carried out by setting a cutoff with tunable $U_j$ and then selecting the value  such that $P_j(R_0,\, 0) \sim 1/2$. As the chain is taken homogeneous and the $j-th$ base pair has been chosen arbitrarily, the selected value holds for all base pairs in the chain, i.e. $U_j \equiv \bar{U}$. Generally, for heterogeneous sequences, one can apply the method to derive a set of site dependent cutoffs.
Thus, the maximum amplitude of the base pair fluctuations is physically related to the set of model potential parameters and to the specific twist conformation for the molecule. The results for a homogeneous fragment of $21$ base pairs are displayed  in Fig.~\ref{fig:2}: the time dependent probability is plotted versus time for different twist conformations defined by five values of the helical repeat $h$. As reported in the inset, the $\bar{U}$ value, that satisfies the zero time probability benchmark, gets larger if the helix unwinds. A physically plausible result.

{From Fig.~\ref{fig:2} we observe that, by taking a molecule with a twist conformation defined by $h=\,10.5$ (this corresponds to the standard room temperature helical repeat for kilo base long DNA \cite{duguet93}), the calculated dimensionless cutoff is $\bar{U} =\,5.2$.  This value, in turn, yields an upper bound $\Lambda_{j}(T)=\,1.08 \AA$ for the first Fourier component in Eq.~(\ref{eq:03b}) which, using Eq.~(\ref{eq:02a}),  leads to a meaningful estimate of $\sim 2.2 \AA$ for the largest pair mates separation in the $j-th$ base pair, that is $\sim 10 \%$ of the average helix diameter. 
In fact, a number of works has shown that such a  fluctuation represents a reasonable estimate for the value above which hydrogen bonds are broken \cite{bax01,pey08,khali22}.
}

Interestingly, similar values for the fluctuation cutoff have been derived by solving analytically the 1D PB model in the thermodynamic limit \cite{io24}. Numerical values for the fluctuation cutoff  as a function of the potential parameters, have been calculated also for the 1D PBD model, see ref.\cite{io20}.

\section*{4.  \, Results and Discussion}

\renewcommand{\theequation}{4.\arabic{equation}}
\setcounter{equation}{0}

After settling the cutoff issue, we apply the statistical model to analyze the interplay between form and helical conformation of nucleic acids, performing a quantitative analysis of their twist-stretch properties. Moreover, we test the model by evaluating the stretching of a DNA chain that is forced to flow through a cylindrical channel and the cyclization probability of linear DNA fragments at short length scales. While the calculations reported hereafter are performed at room temperature, the model can be also extended to study the temperature effects on the flexibility properties, a topic of current interest \cite{io18,shiben23}. 

\subsection*{4.A   \,Twist - Stretch relations}

Despite their apparent similarities, ds-RNA and ds-DNA display distinctive helical structures in physiological conditions which determine different elastic properties
and responses to external perturbations \cite{gonz13,oroz10}. For instance, the A-form ds-RNA has a larger bending rigidity that is, a longer persistence length than the B-form ds-DNA. This follows from the fact that the A-form molecule \textit{i)} is broader and therefore less bendable, \textit{ii)} has a shorter distance between adjacent phosphate groups along the strands hence, the negative charge density and the ensuing electrostatic repulsion between strand segments are higher. In fact, at high salt concentrations the electrostatic repulsion is largely screened and the bending persistence lengths of ds-RNA and ds-DNA converge towards similar values \cite{tan24}. 

Instead, a striking difference between A-form and B-form molecules is seen in magnetic tweezers experiments on kilo-base long molecules: ds-RNA untwists whereas ds-DNA over-twists upon stretching \cite{busta06,dekker14}.

To address this issue and derive the twist-stretch relations in the framework of our model,  one has to modify the Hamiltonian in Eq.~(\ref{eq:01}) by adding the term  $- F_{ex} d_S  \cos\bigl( \phi_n \bigr)$ whereby the load $F_{ex}$ is applied along the molecular axis and uniformly acts on all nucleotides in the chain. 
The general rise distance $d_S$ is defined in Fig.~\ref{fig:1}(c). In the absence of tilt and slide, $d_S \equiv d$.

Next, using a previously developed computational method \cite{io17}, we assume a range of $h$ values as input parameters and, for each $h$, we calculate the average helical repeat $< h >$ by performing integrations over the ensemble of base pair configurations contained in the partition function for the molecule subjected to the load. For each twist conformation, defined by $< h >$, also the free energy is computed.
Accordingly, by minimizing the free energy over a set of possible twist conformations, one selects the equilibrium helical repeat, $< h >_{{*}}$, for a given load. 
The twist profiles ($< h >_{{*}}$ versus $F_{ex}$) are derived by varying the force in the low to intermediate pico-Newton range in which the molecules do not overstretch \cite{busta06}. 

\begin{figure}
\includegraphics[height=8.0cm,width=8.0cm,angle=-90]{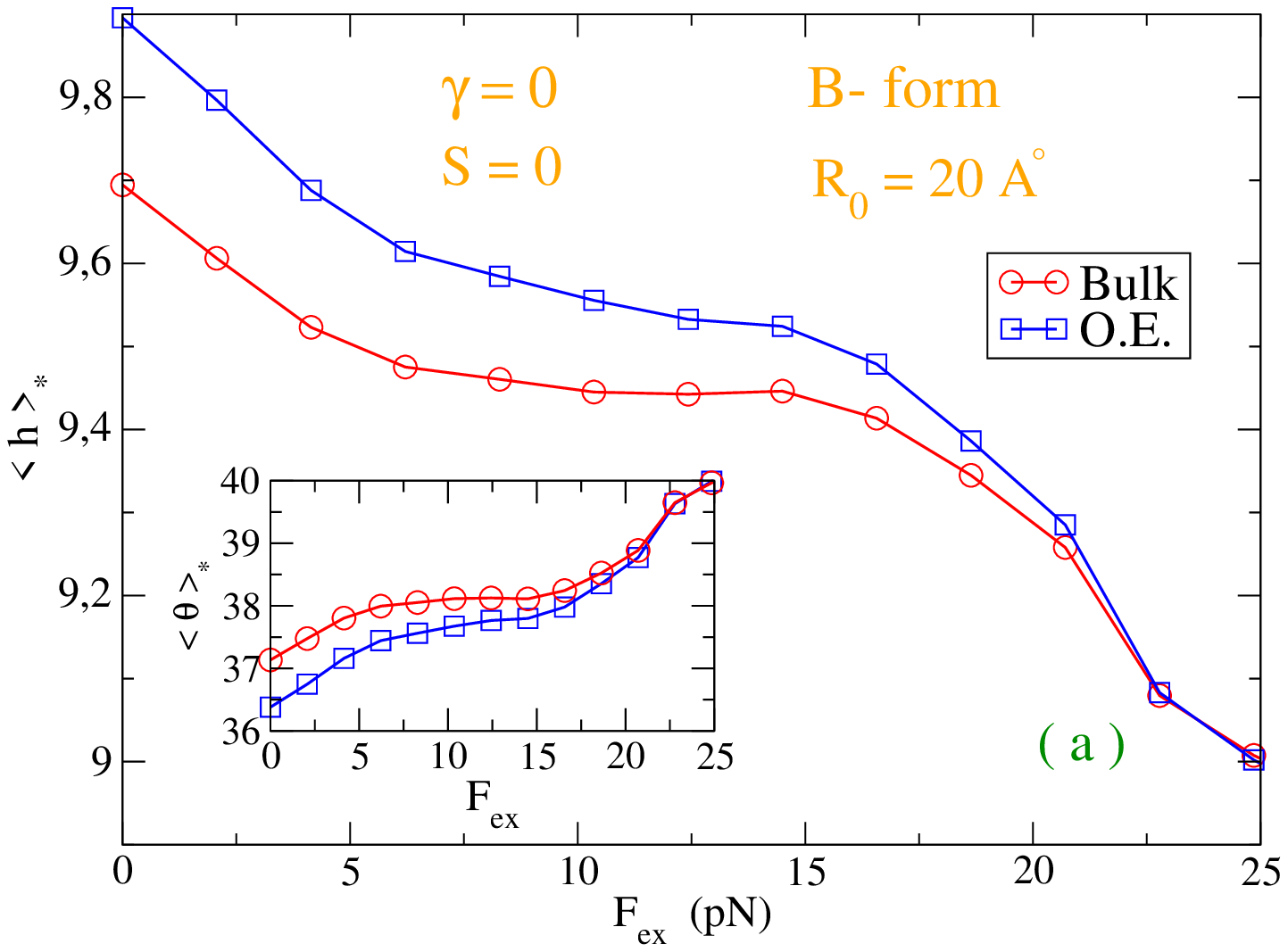}
\includegraphics[height=8.0cm,width=8.0cm,angle=-90]{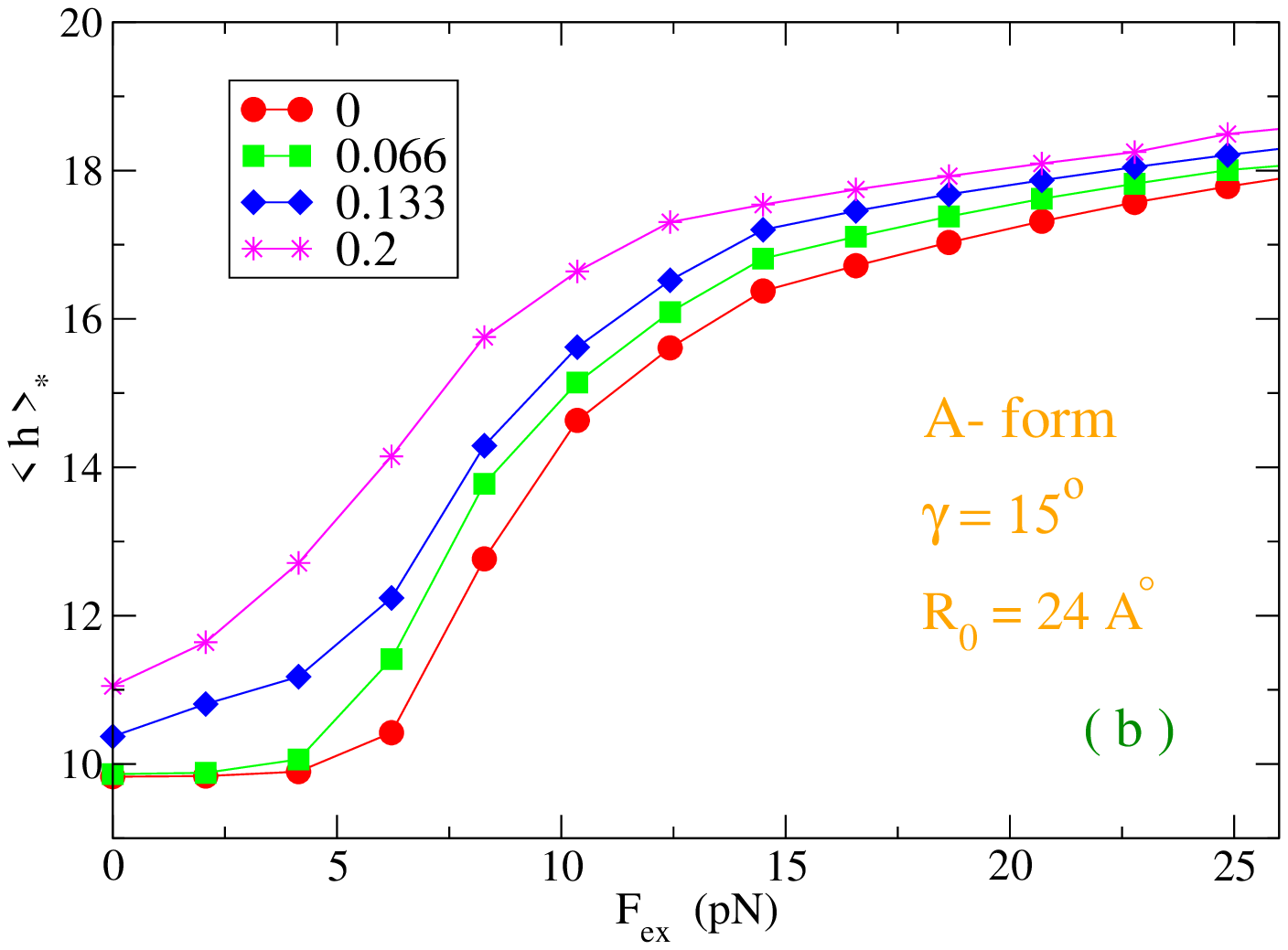}
\caption{\label{fig:3}(Color online) Reprinted  from ref.\cite{io23}. (a)  Equilibrium helical repeat versus stretching force for the B- form helix. The molecule is considered both with (Open Ends) and without (Bulk) terminal base pairs.  The inset shows the average twist angle for both cases.  
(b)  Equilibrium helical repeat versus stretching force for the A- form helix.  Four dimensionless $|S| / d$ ratios are assumed.
}
\end{figure}

The results are shown in Fig.~\ref{fig:3} for the standard B-form of DNA and for the A-form of RNA. Average $R_0$ values for the two forms are taken as input parameters.
To emphasize the role of fraying effects, Fig.~\ref{fig:3}(a) displays the $< h >_{{*}}$ versus the applied load in two cases: i) the full open ends (O.E.) chain made of $N-1$ dimers and ii) the bulk of the chain made of $N-3$ dimers. In the former case, $< h >_{{*}}$ is significantly larger signaling that the terminal base pairs strongly affect the overall helix untwisting thus yielding an enhanced flexibility. This holds both in the absence of loads and for moderate loads up to about $20 \,pN$ whereas, for larger $F_{ex}$ which cause a more pronounced over-twisting \cite{shiben24}, the chain end effects progressively vanish. 

For the A-form, see Fig.~\ref{fig:3}(b), the base pair inclination with respect to the helical axis is taken as $\gamma =\, 15^{o}$ in line with X-ray diffraction data and molecular dynamics simulations \cite{zachar15}. Both the zero slide case and three cases with finite $|S| / d$ values are considered.  
All plots show that $< h >_{{*}}$ grows under the effect of the stretching load revealing that the base pair inclination is the primary cause of the helix untwisting. The effect of the slide is superimposed to that of $\gamma$: by increasing $|S|$, the helix untwisting is larger for all $F_{ex}$ suggesting that the structural features of the dimers determine the overall torsional properties of the chain.

At this stage, it is fair to point out a limitation of the mesoscopic approach. Although our model has assumed a point-like representation for the nucleotides, we remark that the nucleotide has a complex structure with the sugar ring providing the flexible link between nitrogenous base and phosphorous group. Fundamentally, the structural difference between ds-DNA and ds-RNA  is ascribed to the different sugar pucker conformations of the molecules which, in turn, are related to the absence (in DNA) / presence (in RNA) of the hydroxil group in the sugar ring \cite{dick83}. Accordingly, a coarse grained representation as that of Eq.~(\ref{eq:01}) cannot capture the ultimate origin of the opposite behavior envisaged by the twist-stretch relations provided that the latter has to be traced back to the atomic scale.

\subsection*{4.B  \, Stretching DNA in a channel}

\begin{figure}
\includegraphics[height=8.0cm,width=8.0cm,angle=-90]{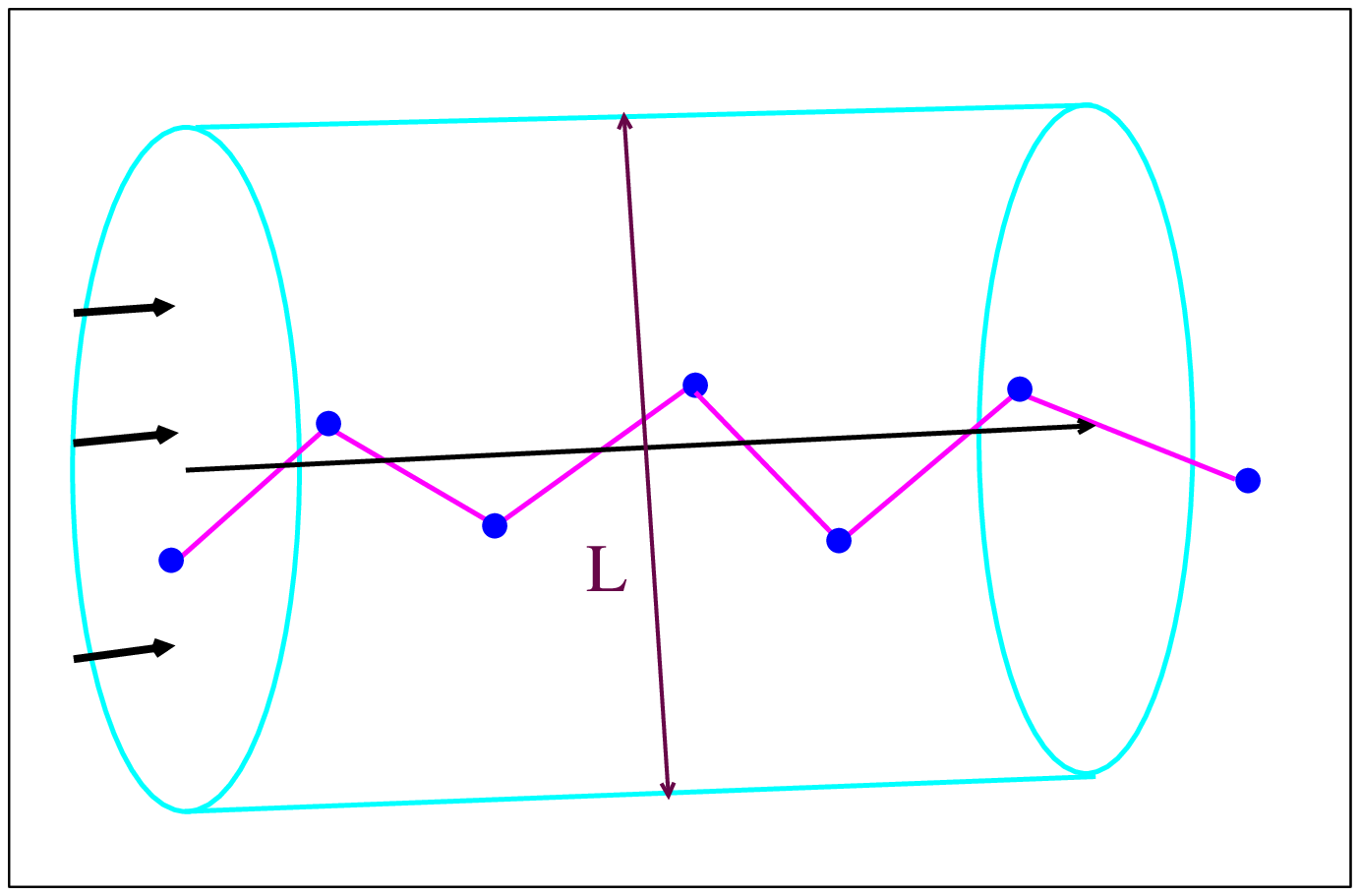}
\caption{\label{fig:4}(Color online)  Reprinted  from ref.\cite{io20b}.
Schematic for the helical chain of Fig.~\ref{fig:1}(a) flowing through a cylindrical channel of variable diameter $L$. 
}
\end{figure}

Next, we focus on a system that models the imaging technique routinely used to stretch the DNA molecules with the purpose to map the genome \cite{kwok12}. The system is schematized by a coiled DNA molecule which is funneled through a nanochannels array, whereby the channel is a cylinder whose tunable diameter $L$, larger than the helix diameter, uniformly confines the base pair fluctuations as shown in Fig.~\ref{fig:4}.

This system can be modeled by a Hamiltonian which is the sum of \textit{i)} the mesoscopic Hamiltonian for the ds-DNA molecule in Eq.~(\ref{eq:01}) and \textit{ii)} a hard-wall potential due to the confining channel ($V_{ch}$) on the radial base pair fluctuations \cite{io20b}. Its expression reads:

\begin{eqnarray}
V_{ch}(r_n)  =\,
\left\{
\begin{matrix}
\chi   \cdot \bigl| |r_n| - R_0 - \delta(L) \bigr|^{-1}  \hskip 0.6cm          |r_n| - R_0 < \delta(L) \\ 
\infty                       \hskip 4.0cm          |r_n| - R_0 \geq  \delta(L) 
\end{matrix} 
\right .  
\label{eq:04}
\end{eqnarray}

$\chi $ and $\delta(L)$ are adjustable parameters weighing strength and range of the force exerted by the cylinder walls on the base pairs motion. 
$V_{ch}(r_n)$ accounts for the physical fact that, by narrowing the channel, the phase space available to the base pair fluctuations is progressively shrunk and  their statistical effect is accordingly reduced. While $\delta(L)$ may not coincide with the diameter $L$, the relation $\, \delta(L) \propto L \,$ should hold.

On general grounds the DNA molecule, subjected to the stretching in a confining channel, should modify its helical conformation. To account for this correlation, the path integral computational method  calculates the average molecule size as a function of the average helical twist. With reference to Fig.~\ref{fig:1}(a), the ensemble averaged end-to-end distance is\,  $< R_{e-e} > =\, \bigl < \bigl| \sum_{n=2}^{N}  \overline{d_{n,n-1}} \bigr| \bigr > $ for the average twist conformation defined by \, $< h >=\,{2\pi N}/ {< \theta_N >}$, with $\theta_N$ being the accumulated twist angle for the last base pair in the chain. By minimizing the free energy, $\, F$, with respect to the set of $< h >$'s values, one selects the equilibrium twist conformation $< h >_{{*}}$ and derives the associated $< R_{e-e} >_{*}$.

\begin{figure}
\includegraphics[height=9.0cm,width=9.0cm,angle=-90]{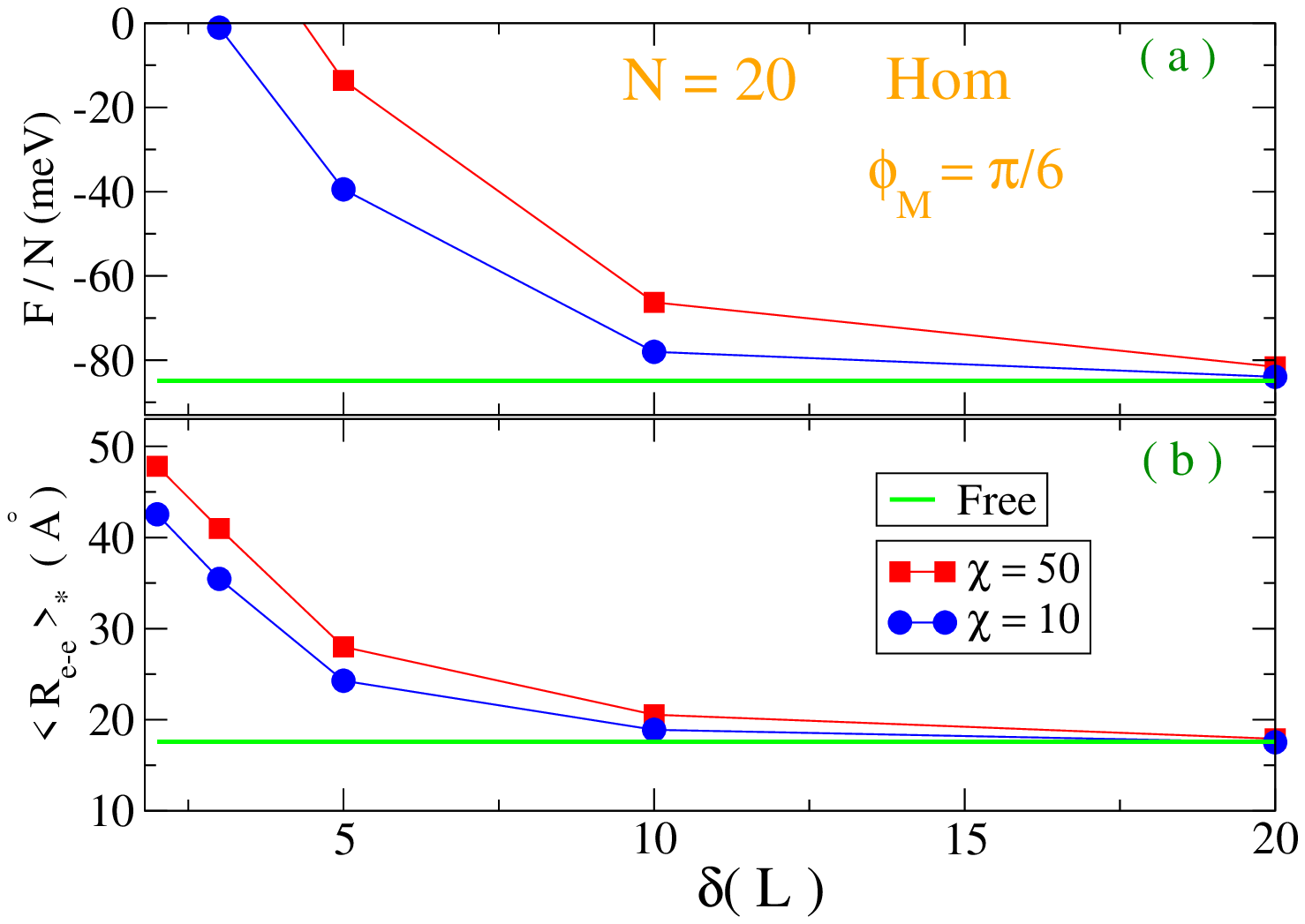}
\caption{\label{fig:5}(Color online)  Reprinted  from ref.\cite{io20b}. (a) Free energy per base pair and (b) equilibrium average end-to-end distance, for a homogeneous chain ($20$ base pairs) in a confining channel.  $\phi_M$ is the maximum bending angle between successive base pair planes (see Fig.~\ref{fig:1}(a)). The effect of the channel diameter is tuned by $\delta(L)$ (units $\AA$). $\chi $ (units $meV \AA^{-1}$) measures the strength of the hard wall potential. The free molecule values are also reported. Both the free energy and the end-to-end distance are calculated for the equilibrium twist conformation. 
}
\end{figure}

As an example, the method is applied to a homogeneous fragment of $N=\,20$  with model parameters suitable to GC base pairs. The free energy per base pair and the $< R_{e-e} >_{*}$ are shown in Fig.~\ref{fig:5} as a function of  $\delta(L)$ for two choices of the interaction strength $\chi $.
Also the results for the free molecule ($V_{ch} \, \equiv 0$) are plotted (green line).  
While for $\delta(L) \sim 20\AA$ the computed values for the confined and free molecule essentially overlap, the free energy significantly increases  for $\delta(L) < 10\AA \,$, markedly for the model with larger $\chi $.  Consistently, the entropic reduction driven by a narrower channel is associated to a substantial stretching of the chain.  The largest calculated end-to-end distance, $< R_{e-e} >_{*} \sim 48 \AA$, corresponds to $\delta(L)=\, 2 \AA$, while even stronger confinements may be achieved by further tuning the channel parameters. This means that the confining cylinder has the effect to straighten the molecule and increase the end-to-end distance by a factor three with respect to the size of the random-coil configuration. For comparison, the contour length of a fully straight oligomer, with $N=\,20$, is $\sim 65 \AA$.

\subsection*{4.C \, Cyclization probability}

Finally, we turn to the problem of the cyclization probability for a set of linear DNA fragments, a key indicator of the molecule flexibility as discussed in the Introduction \cite{marko05,olson06,cherstvy}.

Let's define the $J$- factor as the probability that, given an ensemble of open ends molecules, a fraction of them will bring the chain ends within a given capture volume thus attaining the circular conformation \cite{js50}.   Formally, the $J$- factor is defined by:

\begin{eqnarray}
J=& & \, 8 \pi^2 \frac{Z_{cycle}}{Z_N}  \, , \nonumber
\\
Z_{cycle}=& &\,  \oint Dr_{1} \exp \bigl[- A_a[r_1] \bigr]  \prod_{n=2}^{N}  \int_{- \theta_M }^{\theta_M } d \theta_n  \int_{- \phi_M }^{\phi_M } d \phi_n  \cdot \,  \nonumber 
\\
& & \oint Dr_{n}  \delta^3({\textbf{r} }_{n=\,1} - {\textbf{r} }_{n=\,N}) \exp \bigl[- A_b [r_n, r_{n-1},\phi_n, \theta_n] \bigr] \bigr] \, . \nonumber
\\
\label{eq:05}
\end{eqnarray}

\begin{figure}
\includegraphics[height=8.0cm,width=8.0cm,angle=-90]{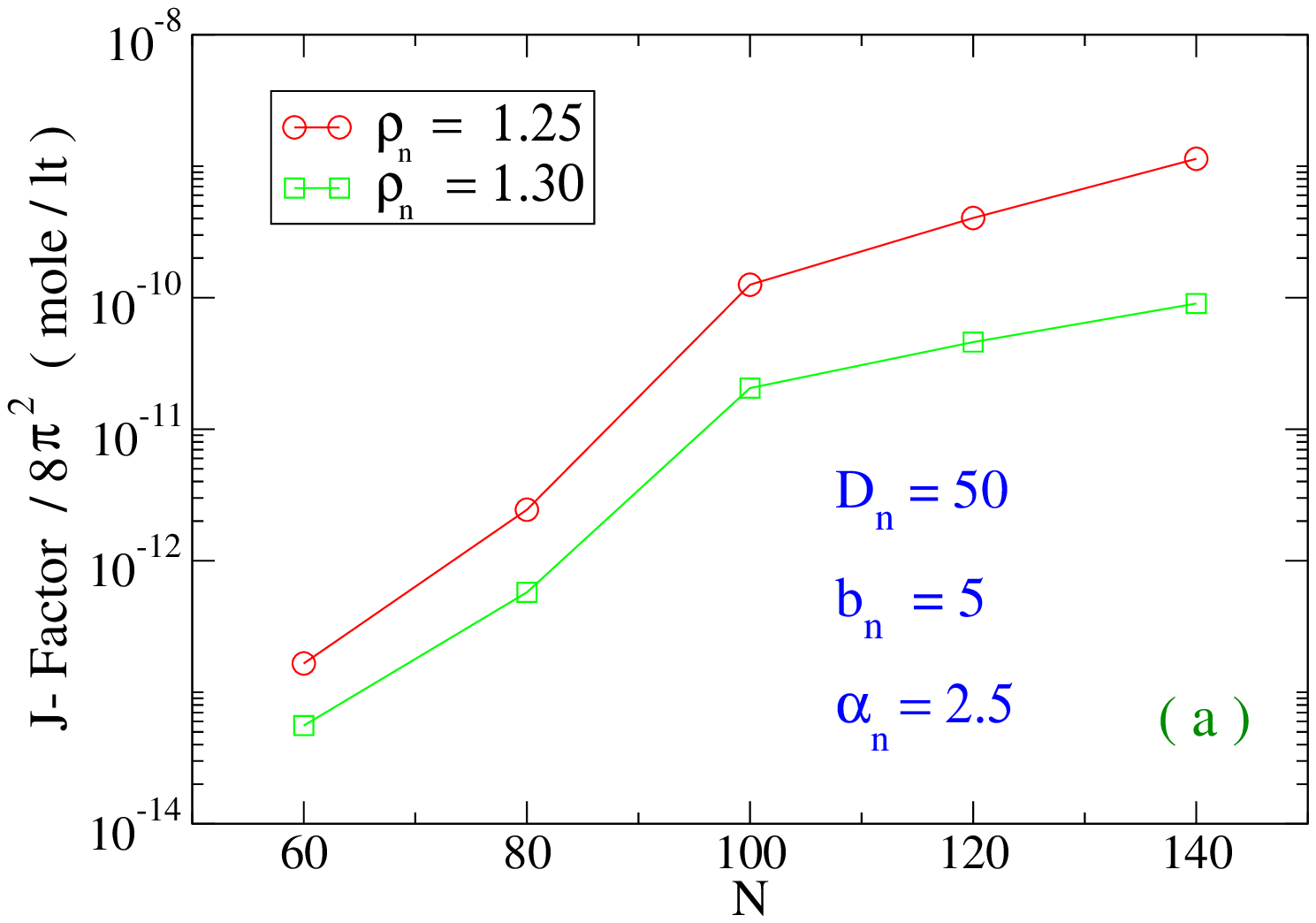}
\includegraphics[height=8.0cm,width=8.0cm,angle=-90]{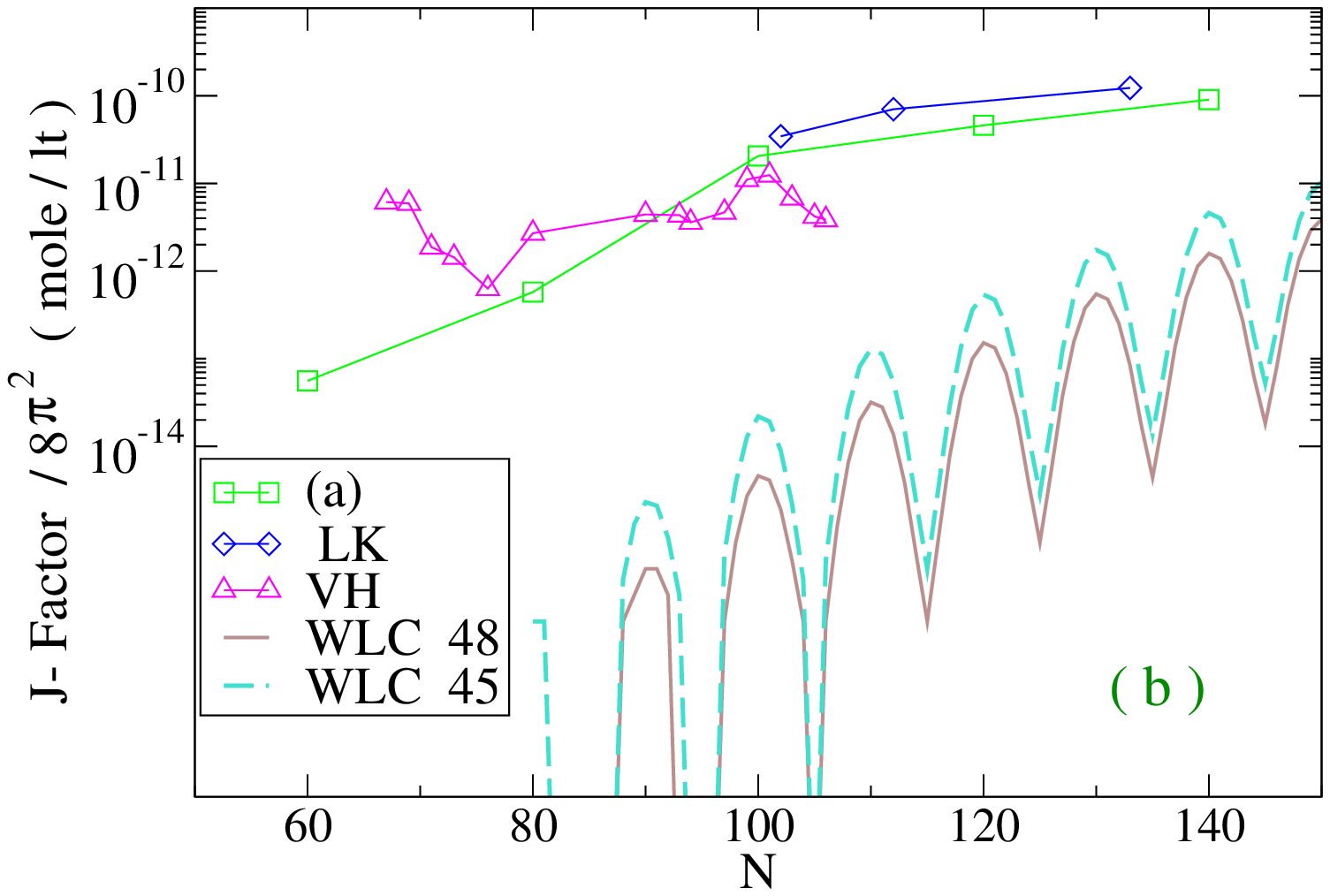}
\caption{\label{fig:6}(Color online)  
Reprinted  from ref.\cite{io11b}. (a) $J$- factor (over $8 \pi^2$) calculated from Eq.~(\ref{eq:05}) for five homogeneous sequences with a number of base pairs $N \in [60 - 140]$.  Dividing Eq.~(\ref{eq:05}) by the Avogadro's number, one gets the units $moles / liter$.   The model parameters are the same for all chains. Two values are assumed for the (dimensionless) nonlinear force constant $\rho _n$. $D_n$ is in units $meV$.  $b_n$ and $\alpha _n$ are in units $\AA^{-1}$.  (b) The green squares in (a) are reported together with the data of ref.\cite{vafa} (VH) and ref.\cite{kim13} (LK). The $J$-factor obtained by the worm-like-chain model (WLC) \cite{shimada}  is plotted for two persistence lengths, $45 \,nm$ and $48 \,nm$.   }
\end{figure}

{$Z_{cycle}$ is the partition function for the molecules whose chain ends have closed into a loop and $Z_{N}$ is the partition function associated to the Hamiltonian in Eq.~(\ref{eq:01}) for open ends molecules. $\theta_M$ and $\phi_M$ are the maximum twisting and bending angles respectively.
}
{$J$ in Eq.~(\ref{eq:05}) is in units of an inverse volume due to the presence of the $\delta^3$-function.  
The numerical program selects the subset of base pair fluctuations consistent with the chain ends constraints given by the $\delta^3$-functions. This is done by imposing that the terminal base pairs in the sequence have, the same $r_n(\tau)$ (for any $\tau$) and equal twisting and bending angles. 
}
{Computing Eq.~(\ref{eq:05}), one notices that the $J$- factor strongly depends on the potential parameters and specifically on the non linear parameters of the stacking potential $\rho _n$ and $\alpha _n$.  Then, the latter can be estimated by comparing the $J$- factor given by the model with the experimentally available looping probability. 
In particular, we have considered the cyclization of molecules yielding a $J$- factor  $\sim  10^{-9}$ mol / liter for $N \sim 100$  as measured by FRET. For this length scale, two independent experiments \cite{vafa,kim13} report consistent values  as shown in Fig.~\ref{fig:6}(b). 
}
{That order of magnitude is assumed as a reference point in order to set a pair of values, e.g. $\alpha _n=\,2.5  \,\AA^{-1}$ and $\rho _n=\,1.3$ although this choice is not unique. The latter values are then taken to calculate the $J$- factor for various chain lengths as plotted in Fig.~\ref{fig:6}(a). 
Also the case with $\rho _n=\,1.25$ is considered to point out the strong dependence of the cyclization probability on the molecule stiffness weighed by the non linear stacking force constant.
}
{While in Sections 4.A and 4.B the equilibrium (ensemble averaged) helical repeat had been determined via free energy minimization, in the current calculation the helical repeat is taken as an input parameter and pinned to the value, $h=\,10$. It follows that  $N / h$ is always an integer for the five considered homogeneous chains whose terminal base pairs can, accordingly, close into a loop without added external twists  \cite{shore81}. Under this conditions, the molecule looping takes place at constant twist conformation as the two strands do not unwind during the process.
For this reason, the well-known oscillating behavior of the $J$- factor (see below) does not appear in Fig.~\ref{fig:6}(a) \cite{popov}.   
}
{The $J$- factor is found to decrease by shortening the fragment size and this trend is all the more evident below $N=\,100$.  
However the chain looping does not drop so sharply as predicted by the WLC model and, e.g. for $N=\,80$, the $J$- factor has still an appreciable value of $\sim 10^{-12}$.  While the results in Fig.~\ref{fig:6}(a) have been obtained for homogeneous fragments, the dependence of the $J$- factor on the
sequence length would not be changed by taking into account heterogeneity effects, nor by assuming different values for the non linear parameters ($\rho _n$, $\alpha _n$). 
}

The results are compared in Fig.~\ref{fig:6}(b) to the cited experimental data obtained for  short oligomers \cite{vafa,kim13}. 
Indeed, some relevant differences exist between the two sets of data indicating the difficulty in deriving the $J$- factors from the measurements  and in comparing the data to the models \cite{towles}.
Nevertheless, the key point on which both experiments concur is that the looping probability is finite and appreciable also at very short molecule lengths and this result is captured by our calculation although a more quantitative comparison between experiments and model could be performed by taking into account also the sequence heterogeneity of the fragments. 

{The remarkable DNA flexibility shown by the plots in Fig.~\ref{fig:6}  can be physically ascribed both to the structure of the stacking potential in Eq.~(\ref{eq:01}) that allows for large bending angles between adjacent base pair planes and to the path integration technique that incorporates in the partition function a broad ensemble of independent base pair fluctuations at any chain site.
}
{In Fig.~\ref{fig:6}(b), we also plot the results calculated using the WLC theory by Shimada and Yamakawa \cite{shimada}  for two values of persistence length: the $J$- factor tends to zero in the low $N$ limit  but finite values can still be obtained for short chains assumed that their persistence length is small, i.e. below $45 nm$.  Measurements of persistence length for short chains are therefore instrumental to establish whether WLC models can predict the looping probabilities at short length scales.
}

Note that the WLC calculation assumes the number of base pairs in the fragment as a continuous variable. Accordingly, the difference between successive peaks in the $J-$ factor provides a measure of $h$. Instead, for $N$ values that are not an integral multiple of the helical repeat, one should add an extra-twist to the DNA helix in order to join the chain ends. Since this extra-twist has an energetic cost, the $J-$ factor decreases. This is the physical origin of the oscillating pattern in the WLC plots. 

Finally, it is pointed out that this study focuses on chains with $ N \sim 100$ and consistently finds that the $J$- factor is a decreasing function of $N$ in such range. On the other hand, if the chain had to be sufficiently long, the random coil behavior would prevail and the looping probability would decrease by increasing $N$. Accordingly, there must be an intermediate range, experimentally found around $N \sim 500$, i.e. about three times the persistence length,  whereby the $J-$ factor attains a maximum \cite{shore83,croth03}. 
Then, extending the calculation in Eq.~(\ref{eq:05}) to molecules with a few hundreds of base pairs would permit \textit{a)}  testing the model in the range in which the ring closure probability displays a non monotonous behavior and \textit{b)} finding the length scale at which the predictions of the mesoscopic Hamiltonian meet those of the WLC model.  This is left for future research.

\section*{5.\, Conclusions}

The mechanical properties of nucleic acids have been extensively characterized over the past decades following the development of a broad range of techniques \cite{chu91,busta00} that allow one to manipulate single molecules and probe their response to external forces. Double stranded (ds) nucleic acids display a remarkable rigidity against bending at the scale of their persistence lengths, albeit maintaining flexibility properties which are instrumental both to their biological functioning and to their applications in the burgeoning field of nanotechnology \cite{saran20,kovaleva24}. Being long and narrow molecules, nucleic acids have been traditionally modeled by bead-spring polymer models whereby each monomer represents hundreds of nucleotides. While this level of coarse graining reduces the myriad of degrees of freedom and permits to simulate the large scale behavior of kilo base long molecules, it is clearly unsuitable to describe the physics of short fragments, let alone of oligomers made of a few tens of nucleotides. At such short scales, mesoscopic models that represent the double helical structure at the level of the nucleotide, provide powerful tools for investigations of the thermodynamical and elastic properties. Here I have reviewed a 3D Hamiltonian model developed over the past few years that contains the fundamental forces stabilizing the molecules and has sufficient detail to account for the bending and twisting fluctuations at specific sites along the double helices. The model has been treated by a statistical method based on the finite temperature path integral formalism that consistently restricts the phase space available to the breathing fluctuations of the base pairs. The theory has been applied to derive some flexibility properties of short fragments of nucleic acids. The computed twist-stretch relations show that the model captures the opposite elastic response of ds-RNA and ds-DNA and explains it in terms of their different helical structures. The calculation of the cyclization probability for ds-DNA chains  shows that fragments with $\sim 100$ base pairs maintain a sizeable bendability in line with the experimental data and at odds with the predictions of the worm-like-chain models of polymer physics. The merits of the mesoscopic approach are pointed out together with its intrinsic limitations and some hints for future developments.

\vskip 0.8cm

\section*{6.\, Declarations}

\textit{Conflict of Interest:}  The author has  no conflicts to disclose.

\vskip 0.5cm
\textit{Funding:} No funding was received for conducting this study.

\vskip 0.5cm
\textit{Data availability:}  The data that support the findings of this study are  available from the corresponding author upon reasonable request.

\end{document}